\documentclass[12pt]{iopart}
\usepackage{iopams}  
\usepackage{graphicx}
\usepackage{hyperref}

\begin{document}

\title[Physical properties of CeGe$_{2-x}$]{Physical properties of CeGe$_{2-x}$ ($x = 0.24$) single crystals}

\author{Sergey L. Bud'ko, Halyna Hodovanets, Alex Panchula\footnote{Currently at First Solar, Inc.},  Ruslan Prozorov, and Paul C. Canfield}

\address{Ames Laboratory, US DOE and Department of Physics and Astronomy, Iowa State University, Ames. IA 50011, USA}

\begin{abstract}
We present data on the anisotropic magnetic properties, heat capacity and transport properties of CeGe$_{2-x}$ ($x = 0.24$) single crystals. The electronic coefficient of the heat capacity, $\gamma \sim 110$ mJ/mol K$^2$, is enhanced; three magnetic transitions, with critical temperatures of $\approx 7$ K,  $\approx 5$ K, and $\approx 4$ K are observed in thermodynamic and transport measurements. The ground state has a small ferromagnetic component along the $c$ - axis. Small applied field, below 10 kOe, is enough to bring the material to an apparent saturated paramagnetic state (with no further metamagnetic transitions up to 55 kOe) with a reduced, below $1 \mu_B$, saturated moment.

\end{abstract}

%Uncomment for PACS numbers title message
% Keywords required only for MST, PB, PMB, PM, JOA, JOB? 
%\vspace{2pc}
%\noindent{\it Keywords}: 
% Uncomment for Submitted to journal title message
\submitto{\JPCM}
% Comment out if separate title page not required
\maketitle

\section{Introduction}

CeGe$_2$ appears to be one of the simple binary compounds that has a benefit of being of a interest for a number of years. Over 50 years ago it was reported to be a ferromagnet with the Curie temperature $T_C \approx 4.5$ K that crystallizes in the orthorhombically distorted ThSi$_2$ structure. \cite{mat58a} The crystal structure of CeGe$_2$ was studied further. \cite{gla59a,gla64a,ian79a,sch91a} It was identified either as a tetragonal $\alpha$-ThSi$_2$ type \cite{gla59a,ian79a}, or as an orthorhombic $\alpha$-GdSi$_2$ type. \cite{gla64a,sch91a}  Magnetization and heat capacity measurements \cite{yas82a} suggested ferromagnetic order with $T_C \approx 7$ K. Similar $T_C \approx 7$ K value and the Kondo temperature $T_K \approx 2.6$ K were reported in \cite{mor85a}. Detailed magnetization study of CeGe$_2$ \cite{lin02a} proposed two magnetic transitions in this material, antiferromagnetic, at $T_N = 7$ K and ferromagnetic, at lower temperature, $T_C = 4.3$ K.

The inconsistencies in determination of the crystal structure in early publications appeared to have a simple explanation. With further studies of the Ce - Ge binary phase diagram it was realized that CeGe$_2$ is not a stoichiometric, line, compound, but has a depleted structure with a width of formation, CeGe$_{2-x}$ with $0.36 \leq x \leq 0.43$, or larger (see below)  \cite{ere71a,ere72a,gok89a,lam90a,lam94a,ven99a}  A transition from high temperature, tetragonal, $\alpha$-ThSi$_2$ type structure to low temperature, orthorhombic,  $\alpha$-GdSi$_2$ type was reported at temperatures between 560\,$^{\circ}$C ($x = 0.36$) and 490\,$^{\circ}$C ($x = 0.43$). \cite{ere71a,ere72a,gok89a}

Within the range of Ge concentrations between 1.71 and 2.0 the orthorhombic crystal structure, and two magnetic transitions, with $T_N \approx 7$ K and $T_C \approx 4.3$ K, virtually independent on Ge concentration, were reported. \cite{zan03a} For lower Ge concentrations, between CeGe$_{1.66}$ and CeGe$_{1.71}$, a tetragonal crystal structure and a single, antiferromagnetic, transition with $T_N \approx 7$ K (and no other magnetic transition down to 0.5 K) were observed. \cite{zan03a} The latter magnetic results were challenged by conclusions in \cite{nak05a}, where two magnetic transitions, one at 6.7 K and another, ferromagnetic, at 5.3 K were described for the  tetragonal CeGe$_{1.66}$. 

Given that majority of the studies, in particular for orthorhombic CeGe$_{2-x}$ with higher Ge concentration,  were performed on polycrystalline samples (as a result the anisotropies of physical properties were not accessed)  and that  the literature data contains apparent inconsistencies, we report a detailed study of the thermodynamic and transport properties of CeGe$_{2-x}$  ($x = 0.24$) single crystals.

\section{Sample characterization and experimental details}

Single crystals of CeGe$_{2-x}$ were grown using the high-temperature solution technique. \cite{can92a,can10a} The constituent elements, with an initial stoichiometry of Ce$_{0.23}$Ge$_{0.77}$, were placed in an alumina crucible and sealed in a fused silica tube under a partial Ar pressure. The ampule was heated up to 1190$^{\circ}$C and then was cooled down to 890\,$^{\circ}$C over $\sim 85$ hours. After that the flux was decanted, meaning that the resulting single crystals were effectively quenched to room temperature from to 890\,$^{\circ}$C over several minutes. The crystals grow as large plates, with the dimensions in excess of $7 \times 7 \times 1$ mm$^3$, with the $c$-axis perpendicular to the plate (see below).

\begin{figure}[tbp]
\begin{center} 
\includegraphics[angle=0,width=120mm]{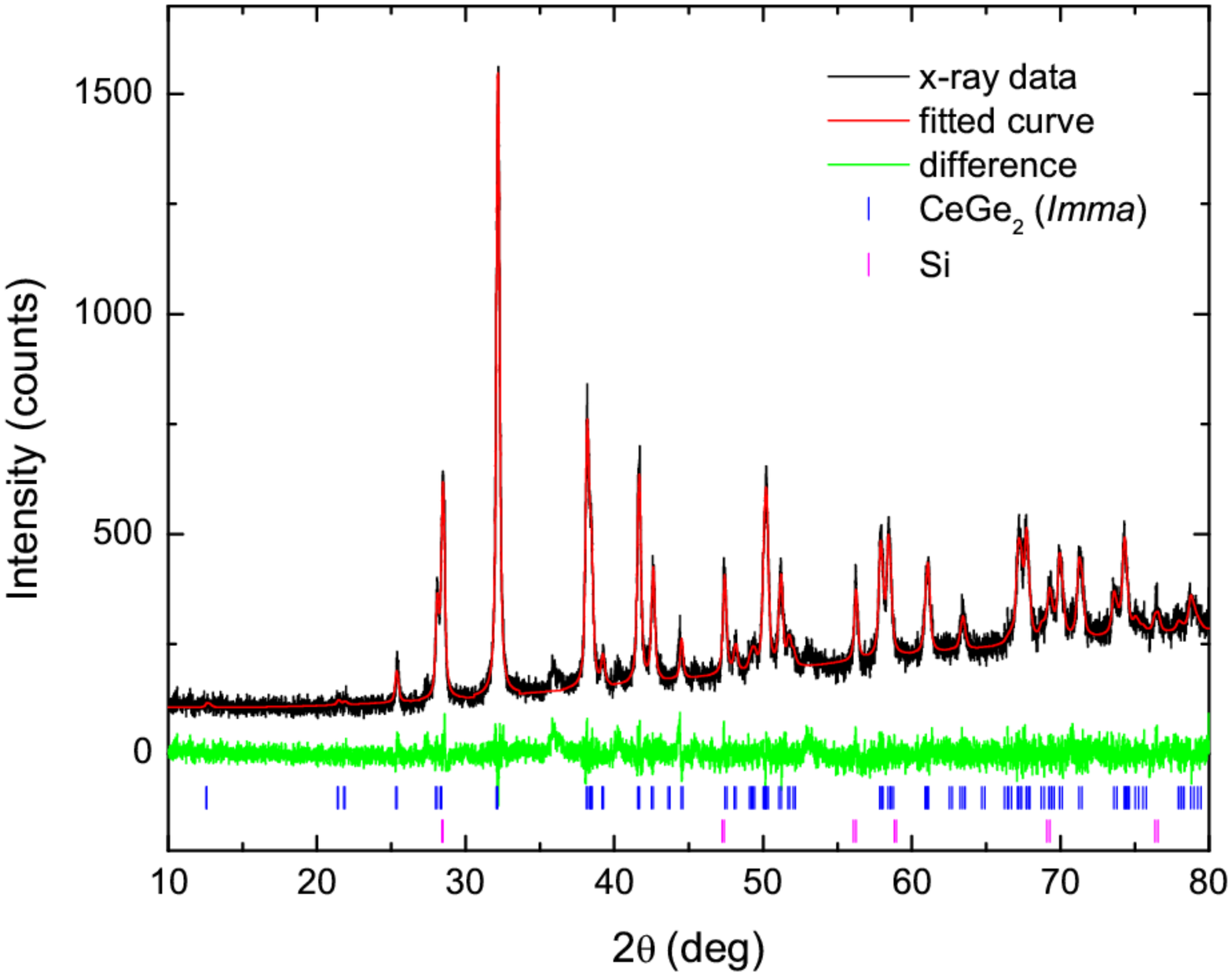}
\end{center}
\caption{(Color online) Powder x-ray diffraction spectrum of ground single-crystal CeGe$_{1.76}$. Note that Si powder was added as a standard. Occupancies were not refined.} \label{F1}
\end{figure}

From earlier work \cite{ere71a,ere72a} and the binary Ce - Ge phase diagram \cite{gok89a} the crystals are expected to have some Ge - deficiency, CeGe$_{2-x}$.   Elemental analysis was performed on these crystals using wavelength-dispersive x-ray spectroscopy (WDS) in the electron probe microanalyzer of a JEOL JXA-8200 electron microprobe. Ce$_2$Fe$_{17}$ and elemental Ge were used as the standards.  The stoichiometry of the samples was determined to be CeGe$_{1.76}$ with the error bar in the value of $x$ of about or less than 0.01. At first glance, it appears that this result contradicts the established width of formation shown in the binary phase diagram.\cite{gok89a} We argue that the spacing between the experimental points in the original papers \cite{ere71a,ere72a} and the error bars present in this work as well as in the original publications allow to remove this apparent contradiction. Given our growth method this result implies that the Ge - rich side of the width of formation should be shifted from Ce$_{38}$Ge$_{62}$ to this composition (Ce$_{36.2}$Ge$_{63.8}$). We will use the notation CeGe$_{1.76}$ for the samples studied in this work.   Room temperature powder x-ray diffraction measurements were taken using Cu-$K_{\alpha1}$ radiation in a Rigaku Miniflex diffractometer. The spectrum was refined using a Rietica software. \cite{hun98a} The x-ray data (Fig. \ref{F1}) are consistent with the sample being orthorhombic with the lattice parameters $a =4.338 \pm 0.009$ \AA,  $b = 4.248 \pm 0.008$ \AA,  and  $c = 14.04 \pm 0.03$ \AA, the values similar to those reported in Refs \cite{gla64a,sch91a,mor85a,ven99a}. It has to be mentioned that a complex superstructure originating from a partial ordering of the Ge vacancies was reported for a sample with a very close Ge-concentration, CeGe$_{1.75}$ \cite{sch91a}. The resolution of our x-ray diffraction data was not sufficient to detect such a superstructure, if present. 

\begin{figure}[tbp]
\begin{center} 
\includegraphics[angle=0,width=120mm]{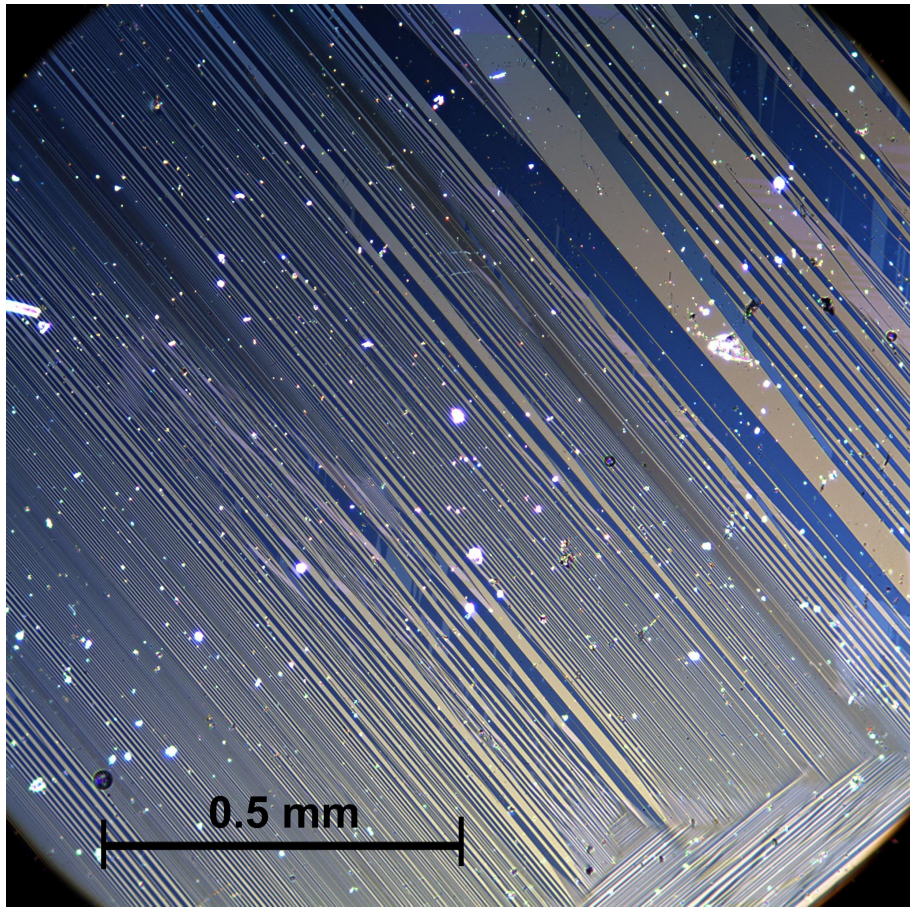}
\end{center}
\caption{(Color online) Optical image of the $ab$ plane of CeGe$_{1.76}$ single crystal at room temperature.} 
\label{F2}
\end{figure}

Figure \ref{F2} shows the optical image of the surface of the crystal taken with optical, polarized light, microscope at room temperature. The very clear domain structure is an evidence of a solid - solid phase transition above room temperature and is fully consistent with the reported tetragonal to orthorhombic structural phase transition near to 500\,$^{\circ}$C. \cite{ere71a,ere72agok89a} The long orthorhombic domains with the very wide range of width, from $\sim 100~\mu$m to  $\sim 1~\mu$m, are clearly seen. The existence of such domains implies that without de-twinning of the crystals measured in-plane properties will represent some average between $a$ and $b$ directions. Additional scattering on the domain boundaries might worsen the in-plane resistivity values.

Anisotropic, temperature dependent, and field dependent dc magnetization was measured using commercial Quantum Design MPMS-5 and MPMS-7 units. In-plane ac resistivity ($I = 1 - 3$ mA, $f = 16$ Hz) was measured using an ACT option of a Quantum Design PPMS-14 instrument. The heat capacity
data  were taken using a hybrid adiabatic relaxation technique of the heat capacity option in a Quantum Design PPMS-14 instrument. Thermoelectric power (TEP)  measurements were carried out by dc, alternating heating (two-heaters-two-thermometers), technique \cite{mun10a} over the temperature range from 2 to 300 K. Thermal expansion and magnetostriction were measured using a capacitive dilatometer constructed of OFHC copper. \cite{sch06a} The dilatometer was mounted in a Quantum Design PPMS-14 instrument and was operated over a temperature range of 1.8 - 300 K in magnetic field up to 140 kOe. Due to aforementioned presence of the orthorhombic domains, the $ab$ dilation was measured along an arbitrary in-plane direction. 

\section{Results and discussion}

\subsection{ Basic physical properties.}

\begin{figure}[tbp]
\begin{center} 
\includegraphics[angle=0,width=120mm]{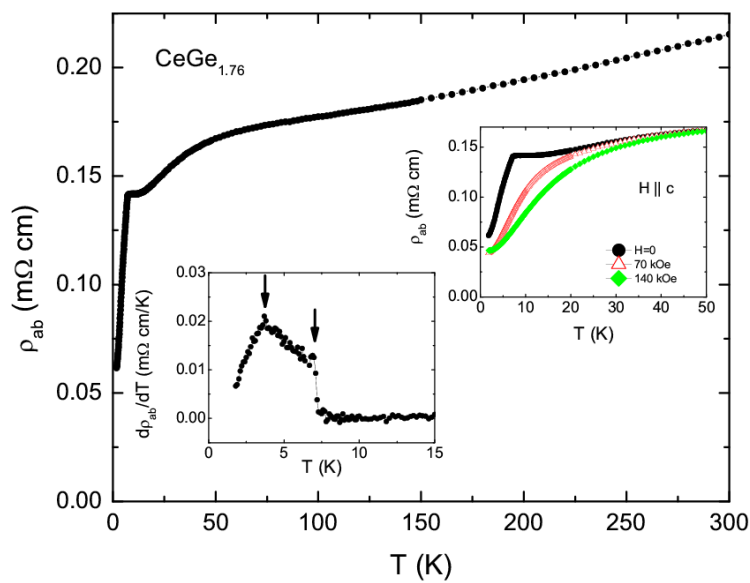}
\end{center}
\caption{(Color online) Temperature-dependent in- plane resistivity of CeGe$_{1.76}$ single crystal. Insets: low temperature resistivity in zero, 70 kOe and 140 kOe magnetic field applied along the $c$-axis (upper right); low temperature part of the derivative $d\rho_{ab}/dT$ with arrows marking the transition temperatures (lower left).} 
\label{F3}
\end{figure}

In-plane, temperature dependent resistivity of the  CeGe$_{1.76}$ crystal is shown in Fig. \ref{F3}. It features a very broad hump at about 50 K and a shallow minimum just above the sharp drop in resistivity below $\sim 7$ K. The residual resistivity ratio, $RRR = \rho_{300 K}/\rho_{1.8 K}$ is approximately 3.5. Such relatively high value of the $RRR$ (compare e.g. with the $RRR$ values of Ni-defficient {\it R}NiGe$_3$, {it R} = rare earth, single crystals \cite{mun10b}) hints at possibility of the ordering of the Ge vacansies suggested earlier, \cite{sch91a} for a sample with a stoichiometry that is essentially CeGe$_{1.75}$. This temperature behavior is similar to  that observed in other heavy fermion or intermediate valence compounds with a low temperature magnetic order. The temperature derivative of the resistivity data, $d\rho_{ab}/dT$, \cite{fis68a} clearly shows two distinct features, at 7.1 K and 3.8 K, suggesting two magnetic transitions in this material. Large magnetic fields suppress the features, and low temperature magnetoresistance is negative. The overall resistivity behavior is similar to what was reported for CeGe$_{1.54}$ \cite{lam94a}, tetragonal CeGe$_{1.66}$ \cite{nak05a} and CeGe$_2$ \cite{mor85a}, despite the differences is stoichometry that  do affect the number and exact temperature of transitions detected in low temperature resistivity.

\begin{figure}[tbp]
\begin{center} 
\includegraphics[angle=0,width=120mm]{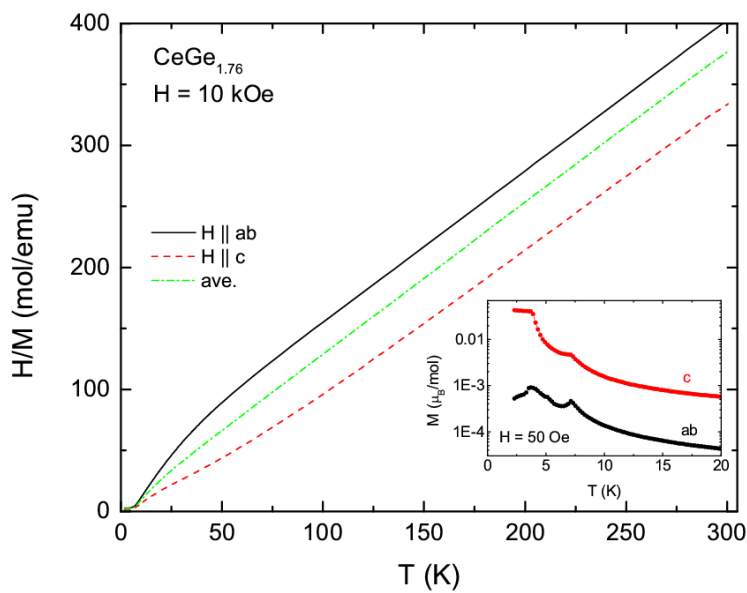}
\end{center}
\caption{(Color online) Anisotropic, inverse, temperature-dependent susceptibility, $1/\chi = H/M$, for CeGe$_{1.76}$ single crystal. The polycrystalline average was calculated as $\chi_{ave} = (2\chi_{ab} + \chi_c)/3$. Inset: low temperature, low field anisotropic magnetization of CeGe$_{1.76}$ single crystal plotted on a semi-log scale.} 
\label{F4}
\end{figure}

\begin{figure}[tbp]
\begin{center} 
\includegraphics[angle=0,width=100mm]{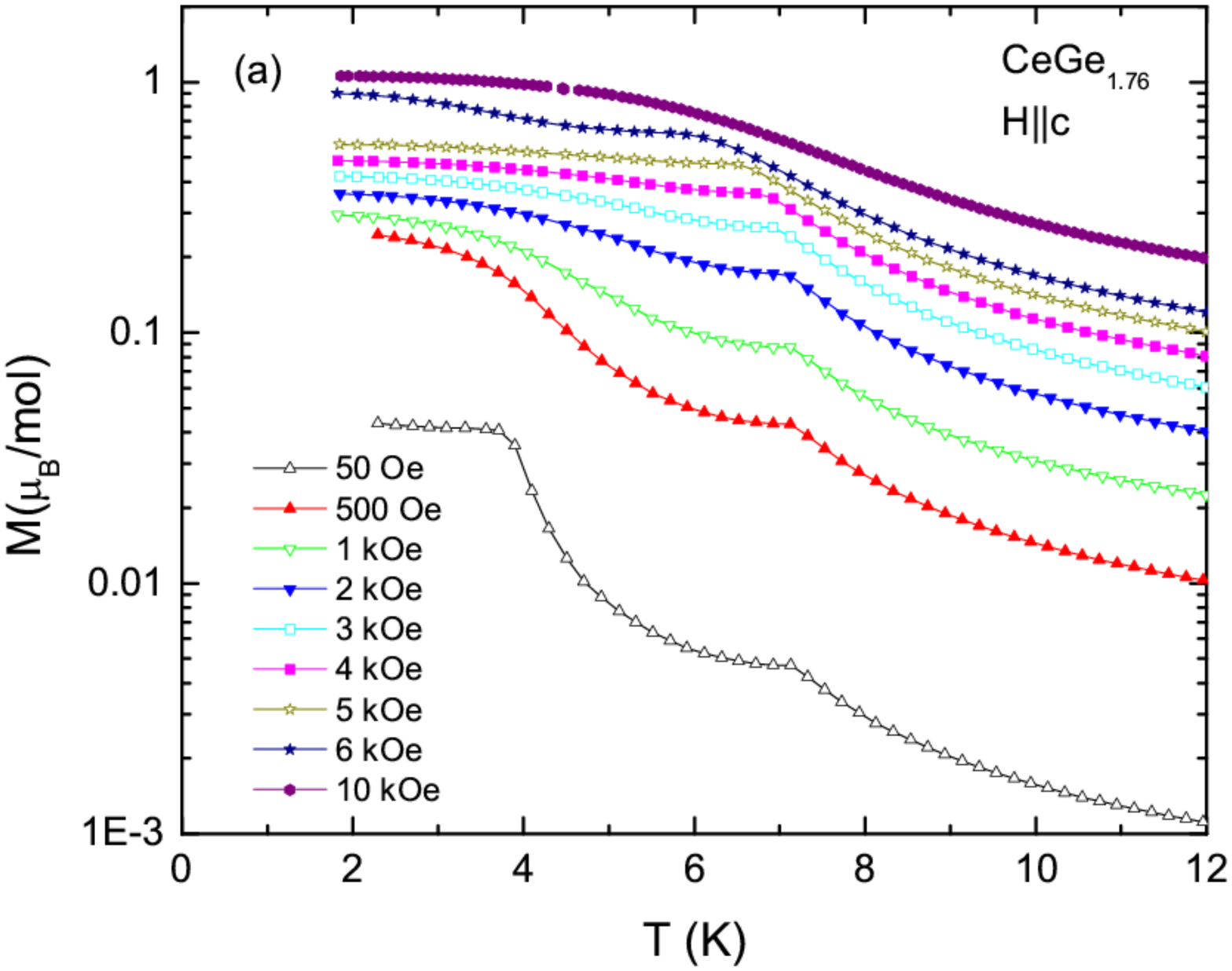}
\includegraphics[angle=0,width=100mm]{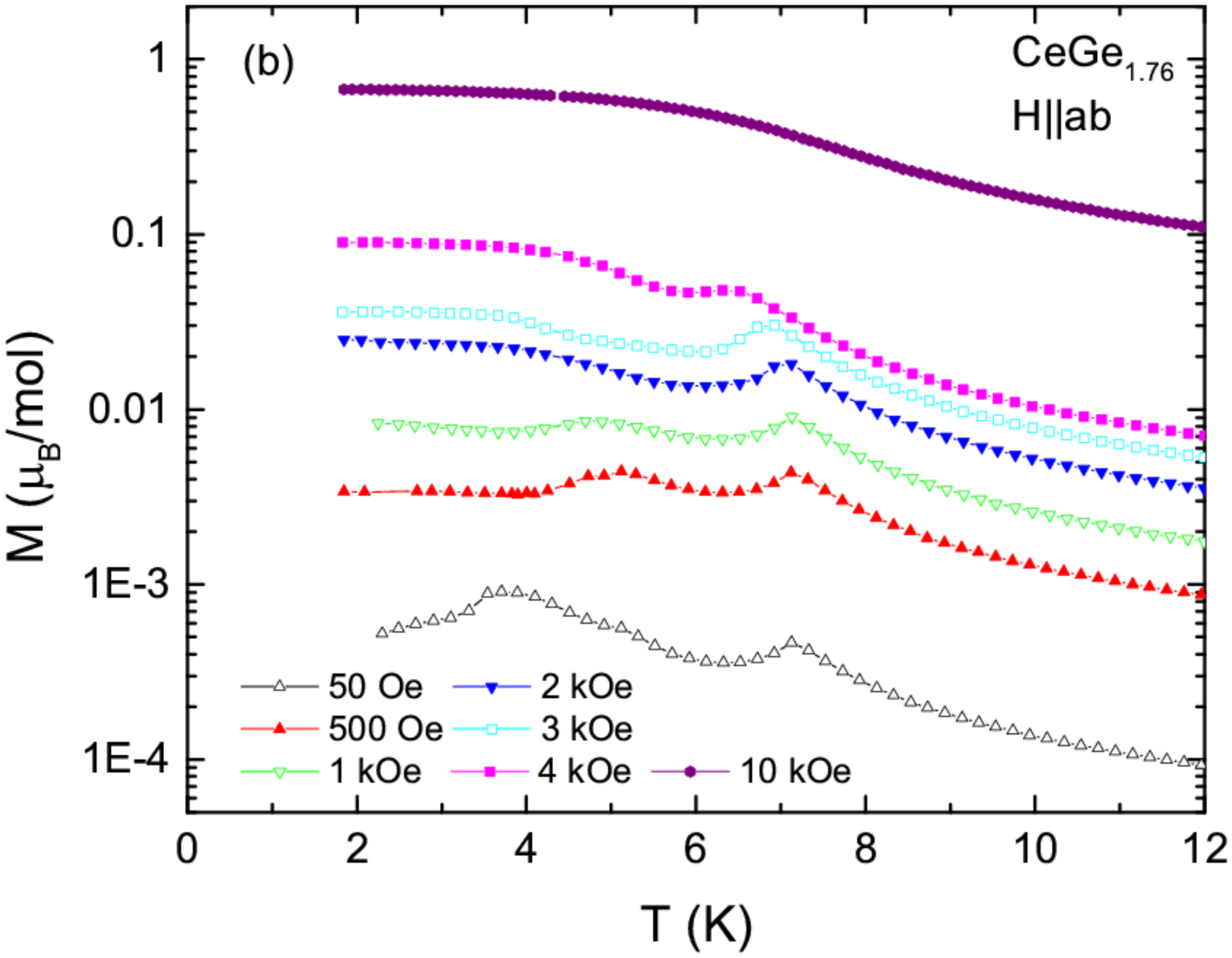}
\end{center}
\caption{(Color online) Low temperature, temperature-dependent magnetization of CeGe$_{1.76}$ measured in several different applied fields for (a) $H \|  c$, and (b) $H \|  ab$, shown in semi-log plots.} 
\label{F10}
\end{figure}

Anisotropic, inverse, temperature-dependent dc susceptibility, $1/\chi = H/M$, measured at 10 kOe is shown in Fig. \ref{F4}. The polycrystalline average was calculated as $\chi_{ave} = (2\chi_{ab} + \chi_c)/3$. The susceptibility is anisotropic with $\chi_{ab} < \chi_c$ in the paramagnetic state. The Curie - Weiss fits of the high temperature (150~K $\leq T \leq$ 300~K) susceptibility data, $\chi = C/(T - \Theta)$ (where $C$ and $\Theta$ are Curie constant and Curie - Weiss temperature, respectively) yield the effective moment (from $\chi_{ave}$) $\mu_{eff} = 2.53 \mu_B$, consistent with the theoretical value for Ce$^{3+}$, and the Curie - Weiss temperatures, $\Theta_{ab} = -25.4$ K, $\Theta_c = 21.9$ K, and  $\Theta_{ave} = -4.4$ K. The signs of the Curie - Weiss  temperatures suggest that antiferromagnetic interactions dominate in the $ab$-plane whereas ferromagnetic ones are predominant along the $c$-axis. The low temperature, low field magnetization data are shown in the inset to Fig. \ref{F4}. These data further confirm magnetic transitions at $\sim 7$ K and $\sim 3.8$ K. In addition, the data for $H \| ab$ suggests one more transition, at $\sim 5$ K (see Fig. \ref{F10} and discussion below).

\begin{figure}[tbp]
\begin{center} 
\includegraphics[angle=0,width=120mm]{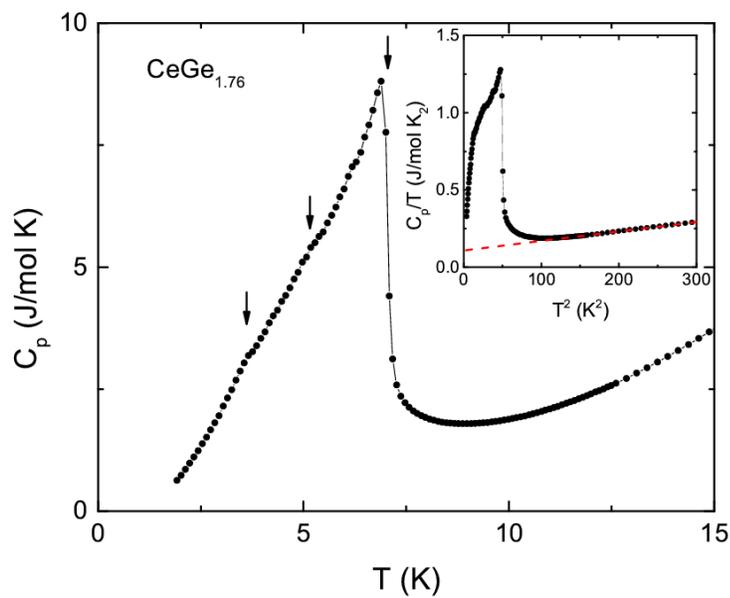}
\end{center}
\caption{ Low temperature heat capacity, $C_p(T)$ of  CeGe$_{1.76}$ single crystal. Arrows point to the features associated with the magnetic transitions. Inset: the same data plotted as $C_p/T$ vs $T^2$. Dashed line - linear extrapolation of $C_p/T$ vs. $T^2$ from above the magnetic transitions. } 
\label{F5}
\end{figure}

The temperature dependent heat capacity (Fig. \ref{F5}) has a clear, distinct feature at $\sim 7.0$ K and somewhat less pronounces anomaly $\sim 3.7$ K (seen as a knee when the data are plotted as $C_p/T$ vs $T^2$, Fig. \ref{F5}, inset). There appear to be other small deviations in $C_p(T)$ behavior in the 3.7 K $<~T~<$ 7 K temperature range. They do not have corresponding anomalies in resistivity, however the one at $\sim 5$ K is consistent with the anomaly in the in-plane low field magnetization, mentioned above. The Sommerfeld coefficient, estimated from the linear fit of $C_p/T$ vs $T^2$ above the transitions, between the  150 K$^2$ and 300 K$^2$ values of $T^2$ (12 - 17 K in the units of $T$) is enhanced, $\gamma \approx 110$ mJ/mol K$^2$. This value of the Sommerfeld coefficient is similar to that reported for CeGe$_2$ and CeGe$_{1.83}$ \cite{lin02a,zan03a} and is noticeably higher than the values reported for  CeGe$_{1.68}$ and  CeGe$_{1.66}$. \cite{zan03a,nak05a} A rough estimate of the magnetic entropy associated with the magnetic transitions (integration up to $T \approx 7.5$ K) gives the value of $S_m \approx 0.75~R\ln2$ .

\begin{figure}[tbp]
\begin{center} 
\includegraphics[angle=0,width=120mm]{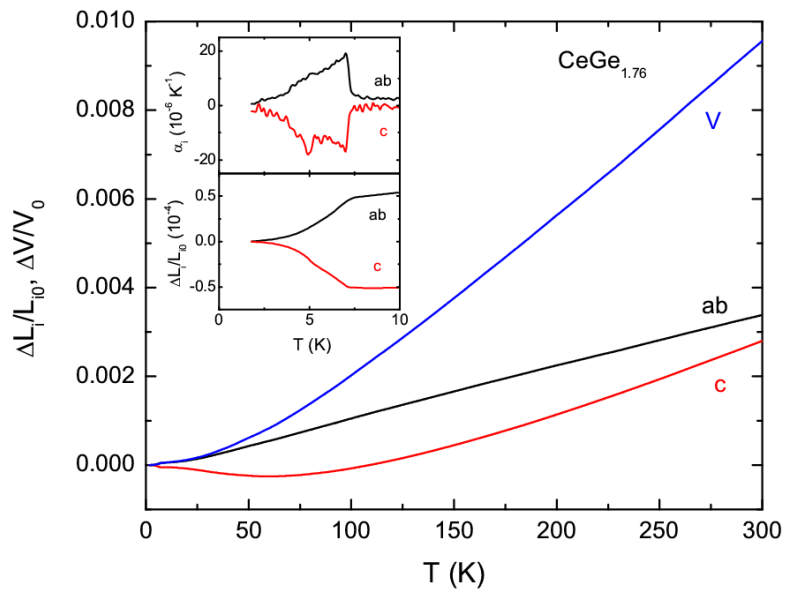}
\end{center}
\caption{(Color online) Temperature-dependent anisotropic, $ab$-plane and $c$-axis, dilation and the volume change of CeGe$_{1.76}$. The data are normalized to the respective values at 1.8 K. Inset: low-temperature $ab$-plane and c-axis dilation and anisotropic thermal expansion coefficients. } 
\label{F6}
\end{figure}

Anisotropic temperature-dependent dilation of  CeGe$_{1.76}$ is shown in Fig. \ref{F6}. The overall change of volume from the base temperature, 1.8 K to 300 K is close to 1\%, that is similar to the change of volume in Cu \cite{kro77a} and is approximately factor of two larger than that of YNi$_2$B$_2$C \cite{bud06a}. In - plane thermal dilation is positive over the whole temperature range and is close to linear in temperature above $\sim 50$ K with the thermal expansion coefficient $\alpha_{ab} \approx (1.1 - 1.2) \times 10^{-5}$ K$^{-1}$. Negative thermal expansion along the $c$-axis is observed in CeGe$_{1.76}$ below $\sim 60$ K with the volume thermal expansion being positive in the whole temperature range. The upper transition is clearly seen at $\approx 7.0$ K in both anisotropic dilation and thermal expansion coefficients (Fig. \ref{F6}, inset). The lower transition is seen as a small step-like feature at  $\approx 3.9$ K in $\alpha_{ab}$.   In $\alpha_c$ a well-defined minimum at  temperature $\approx 4.9$ K  is seen, in agreement with the positions of anomalies in the $M_{ab}(T)$ and the heat capacity. 

Since the magnetic transition at $\sim 7$ K appears to be the second order, we can use our heat capacity and thermal expansion results to evaluate the initial uniaxial and total pressure derivatives of this transition using the Ehrenfest relation,  \cite{bar99a}:  $dT_{crit}/dp_i = \frac{V_m \Delta \alpha_i T_{crit}}{\Delta C_p}$, where $V_m$ is the molar volume ($V_m \approx 3.9 \cdot 10^{-5}$ m$^3$ using the lattice parameters from the X-ray diffraction), $\Delta \alpha_i~ (i = a, c)$ is the jump in the thermal expansion coefficient at the phase transition, and $\Delta C_p$ is the corresponding jump in the heat capacity. Using the experimental data above, for CeGe$_{1.76}$ we found out that in - plane uniaxial pressure derivative is positive, $dT_{crit}/dp_{ab} \approx 0.042$ K/kbar, whereas the $c$-axis uniaxial pressure derivative is  very similar in its absolute value but negative $dT_{crit}/dp_{ab} \approx - 0.040 $ K/kbar. The hydrostatic pressure derivative can be approximated as $dT_{crit}/dP = 2 dT_{crit}/dp_{ab} + dT_{crit}/dp_c \approx 0.044$ K/kbar. So it appears that  CeGe$_{1.76}$ is equally sensitive to uniaxial pressure applied in the $ab$ - plane and along the $c$-axis, however the upper magnetic transition temperature increases when the pressure is applied in the $ab$ - plane and decreases when it is applied along the $c$-axis. The absolute values of the pressure derivatives are fairly small. The absolute value of the uniaxial pressure derivatives are close to those inferred for orthorhombic CeVSb$_3$ \cite{col11a}, however for CeVSb$_3$ all uniaxial pressure derivatives have the same, positive, sign. CeGe$_{1.76}$, like CeVSb$_3$, may be close to a local maximum in ordering temperature and pressures up to 7 GPa may well be able to suppress the magnetic transition to a quantum phase transition.

\begin{figure}[tbp]
\begin{center} 
\includegraphics[angle=0,width=120mm]{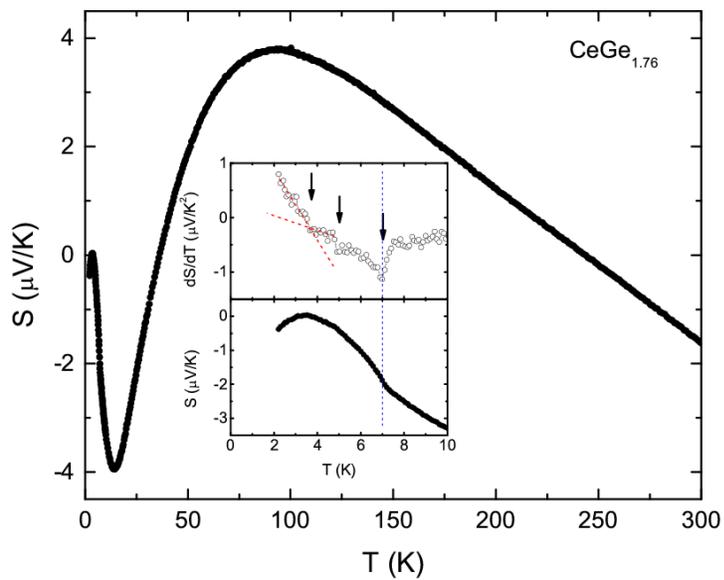}
\end{center}
\caption{(Color online) Temperature-dependent in-plane ($\nabla T \perp c$) TEP, $S(T)$, of CeGe$_{1.76}$.  Inset: low-temperature $S(T)$ and temperature derivative, $dS/dT$. Arrows mark transitions, dashed lines are guides for the eye.} 
\label{F7}
\end{figure}

Temperature-dependent, in-plane, TEP data for CeGe$_{1.76}$ is shown in Fig. \ref{F7}. On a gross level these data are similar to those reported for CeGe$_2$. \cite{jac90a}  The data show two maxima, a broad one at $\sim 90$ K (corresponding to the lowest crystal electric field splitting of the energy levels \cite{lah87a}) and a sharper one, at $\sim 3.5$ K and a minimum at $\sim 14$ K.  Since it is expected that $S(T = 0) = 0$, another minimum, below 2 K is anticipated. Above $\sim 150$ K the TEP is linear in temperature,  indicating that in this temperature region the diffusion TEP is dominant. The TEP changes sign several times, being negative above $\sim 242$ K, positive between $\sim 36$ K and $\sim 242$ K, then negative below $\sim 36$ K with a very short positive excursion between 3.0 K and 3.8 K. Two low temperature transitions, at $\approx 7.0$ and $\approx 3.7$ K, are clearly seen in the $dS/dT$ (Fig. \ref{F7}, inset), the upper one as a sharp minimum and the lower one as a change in slope. There appears to be an additional, step-like feature in the $dS/dT$ at $\approx 5$ K, close to the temperature of small anomalies in the thermodynamic measurements discussed above.

\subsection{Anisotropic $H - T$ phase diagrams}

\begin{figure}[tbp]
\begin{center} 
\includegraphics[angle=0,width=100mm]{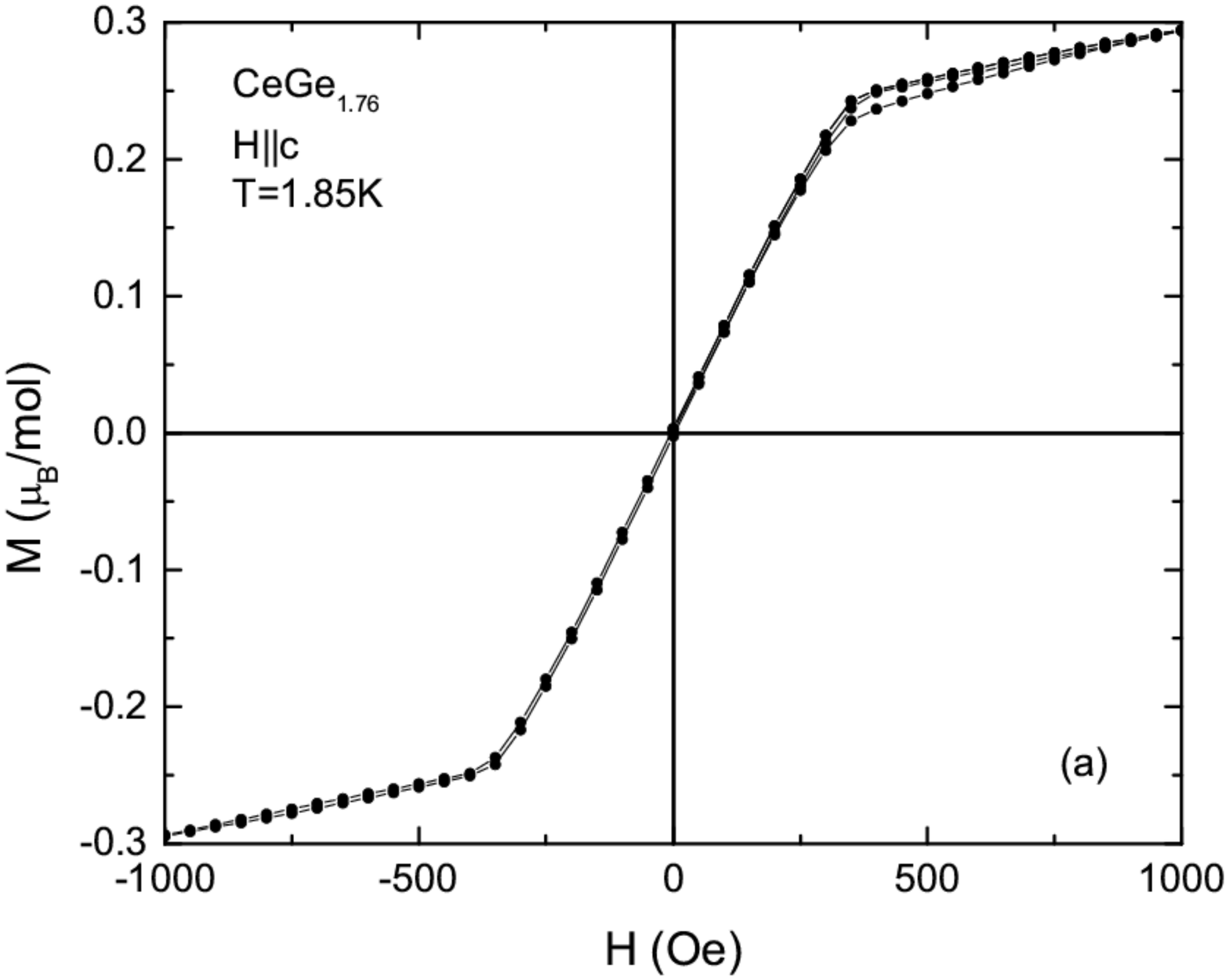}
\includegraphics[angle=0,width=100mm]{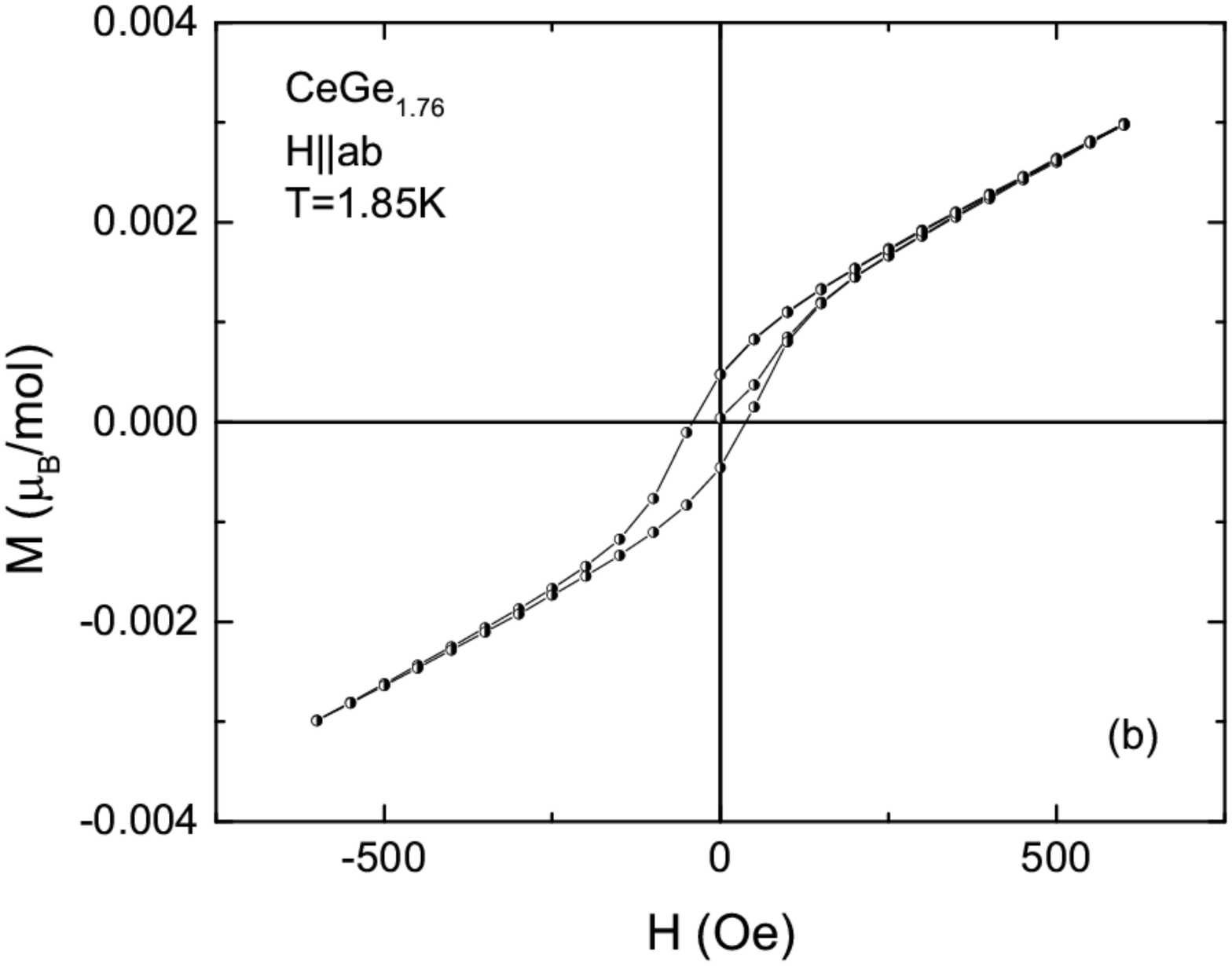}
\end{center}
\caption{Low-field, five-quadrants magnetization loops  of CeGe$_{1.76}$ measured at $T = 1.85$ K for (a) $H \|  c$, and (b) $H \|  ab$.} 
\label{F8}
\end{figure}

Low-field, five-quadrant, anisotropic magnetization loops  for CeGe$_{1.76}$ are shown in Fig. \ref{F8}. These data suggest that there is a ferromagnetic component in the magnetization along the $c$ - axis at 1.85 K, with a rather small, $\approx 0.22~\mu_B$ per cerium, saturated moment and very narrow hysteresis loop. The width of the loop and the saturated moment along the $c$ - axis are smaller than those of stoichiometric and magnetically soft CeAgSb$_2$ \cite{mye99a}. This virtual absence of hysteresis again hints at possible order of vacancies (perhaphs in a CeGe$_{1.75}$, or Ce$_4$Ge$_7$, structure) or at least lack of pinning, even on domain walls.  The apparent low field ferromagnetic loop measured for the magnetic field in the $ab$ plane might be intrinsic, but very possibly is a result of a slight (less than 0.5\,$^{\circ}$) misallignment of the sample. 

\begin{figure}[tbp]
\begin{center} 
\includegraphics[angle=0,width=100mm]{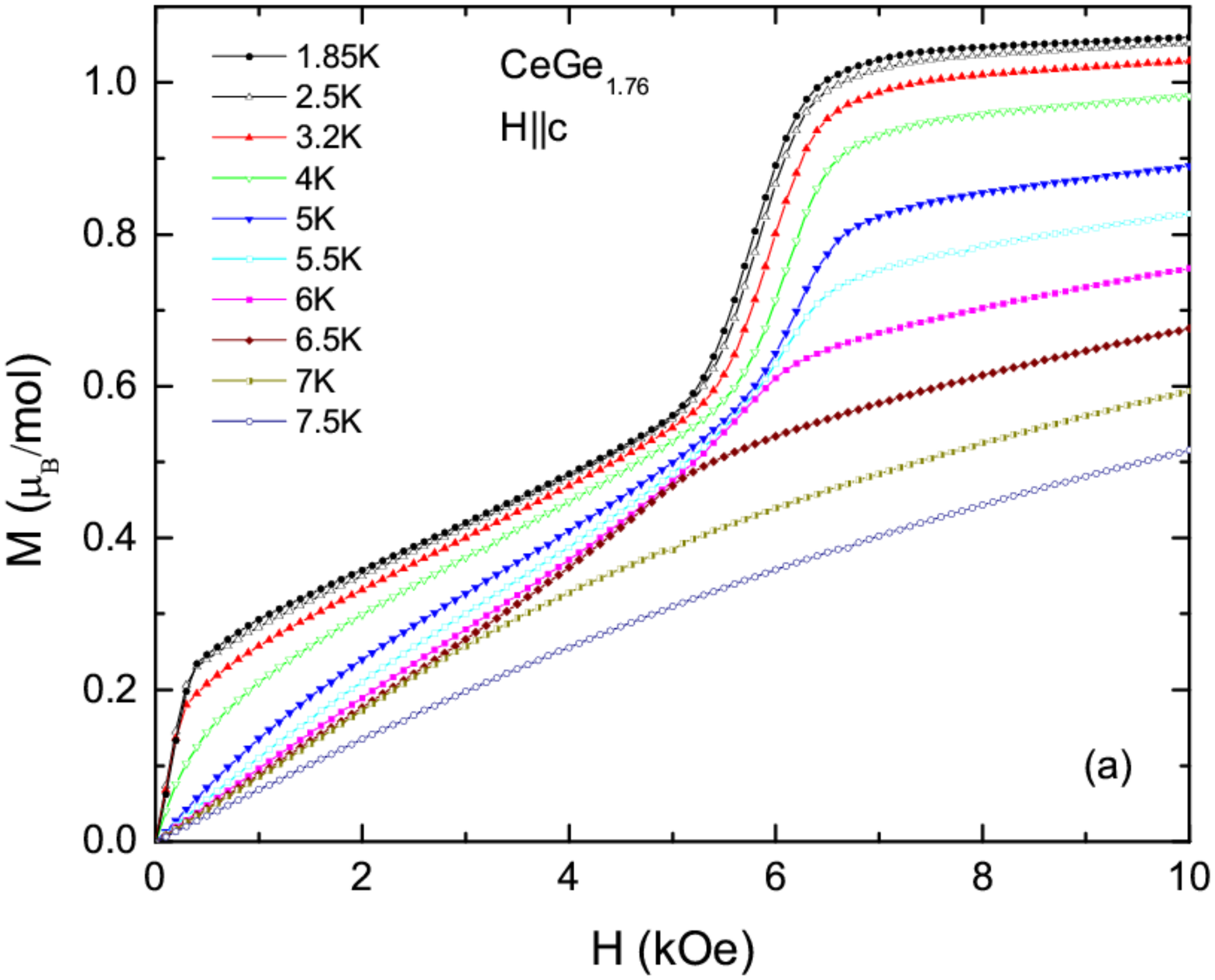}
\includegraphics[angle=0,width=100mm]{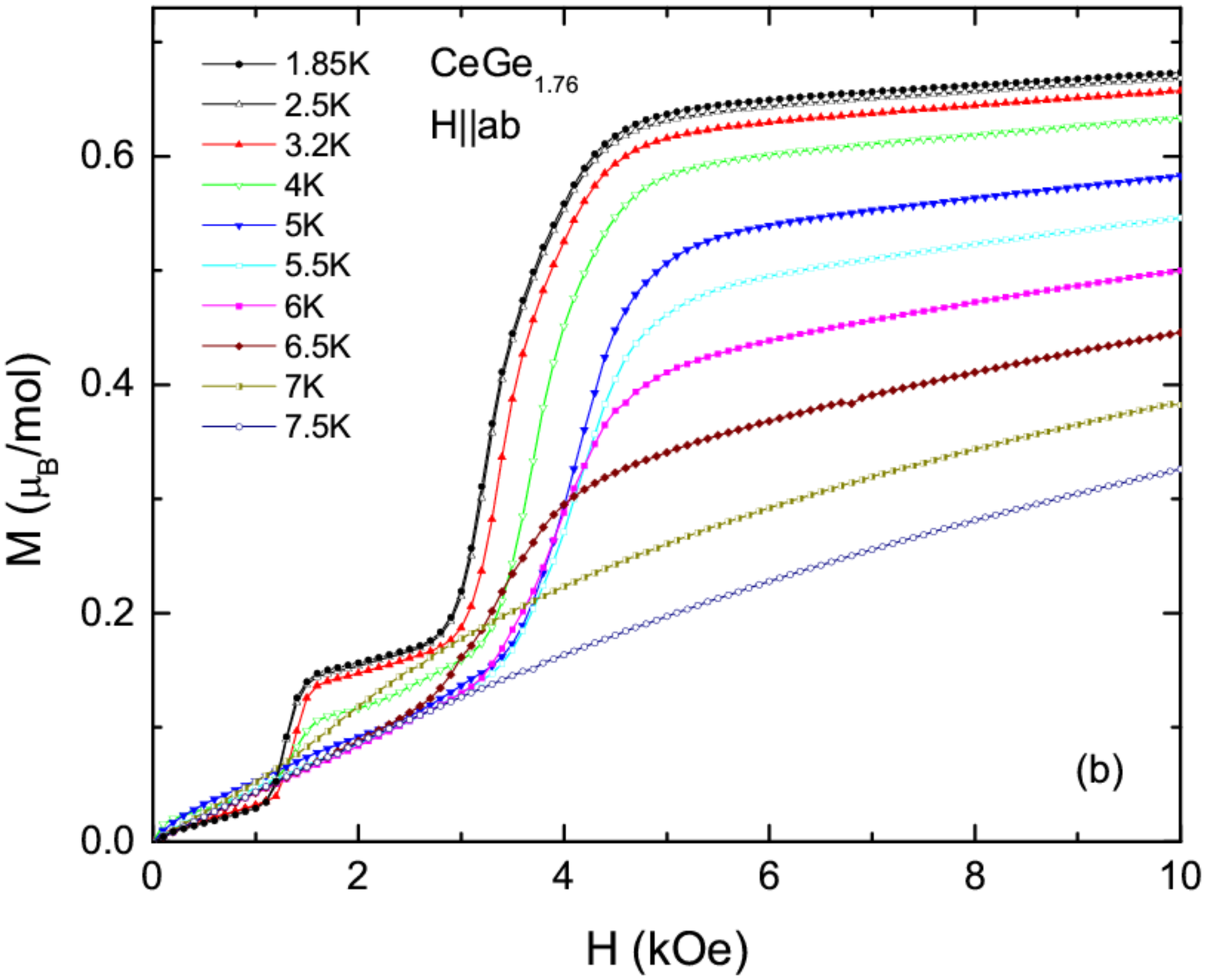}
\end{center}
\caption{(Color online) Field-dependent magnetization of CeGe$_{1.76}$ measured at several different temperatures for (a) $H \|  c$, and (b) $H \|  ab$.} 
\label{F9}
\end{figure}

\begin{figure}[tbp]
\begin{center} 
\includegraphics[angle=0,width=100mm]{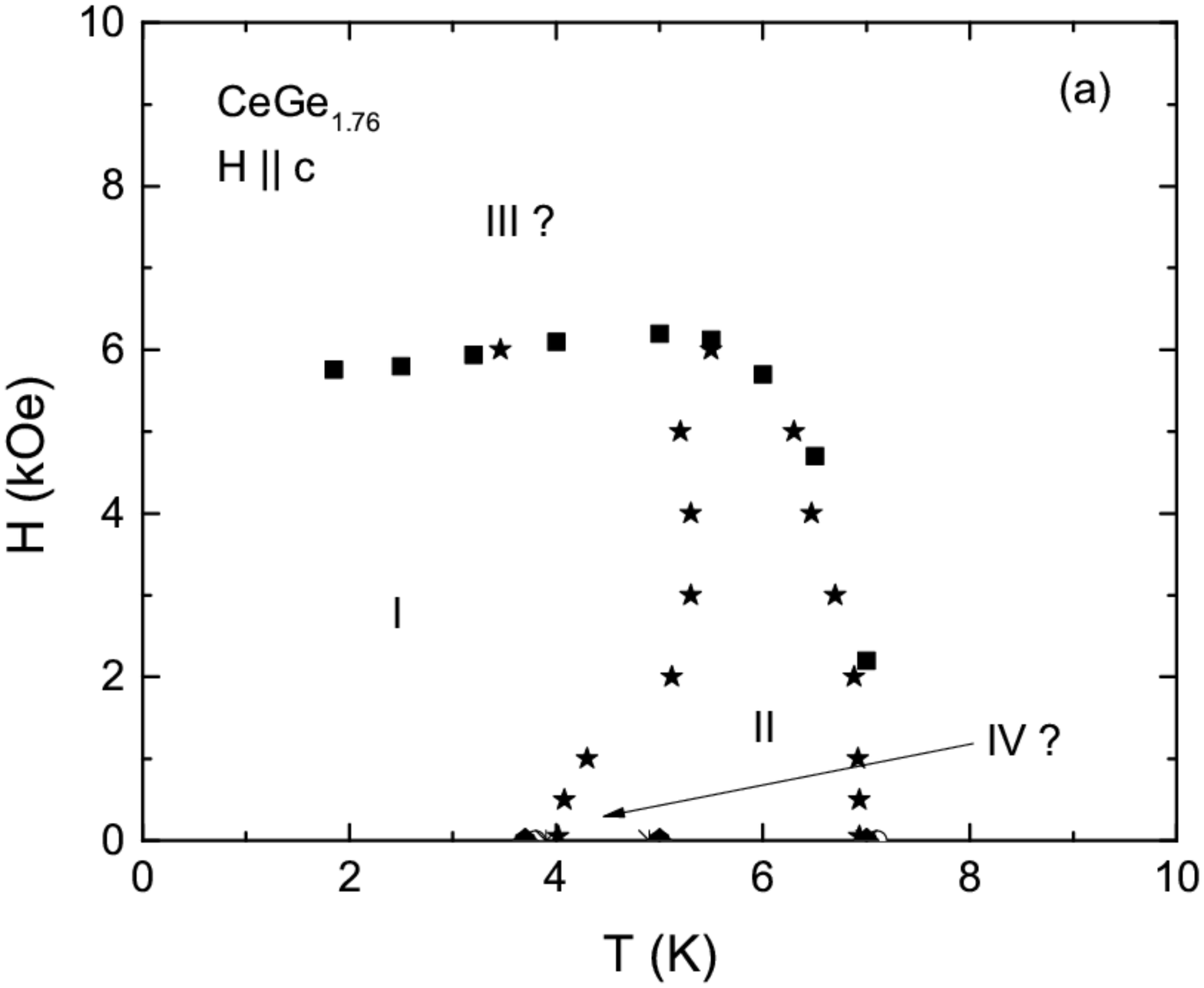}
\includegraphics[angle=0,width=100mm]{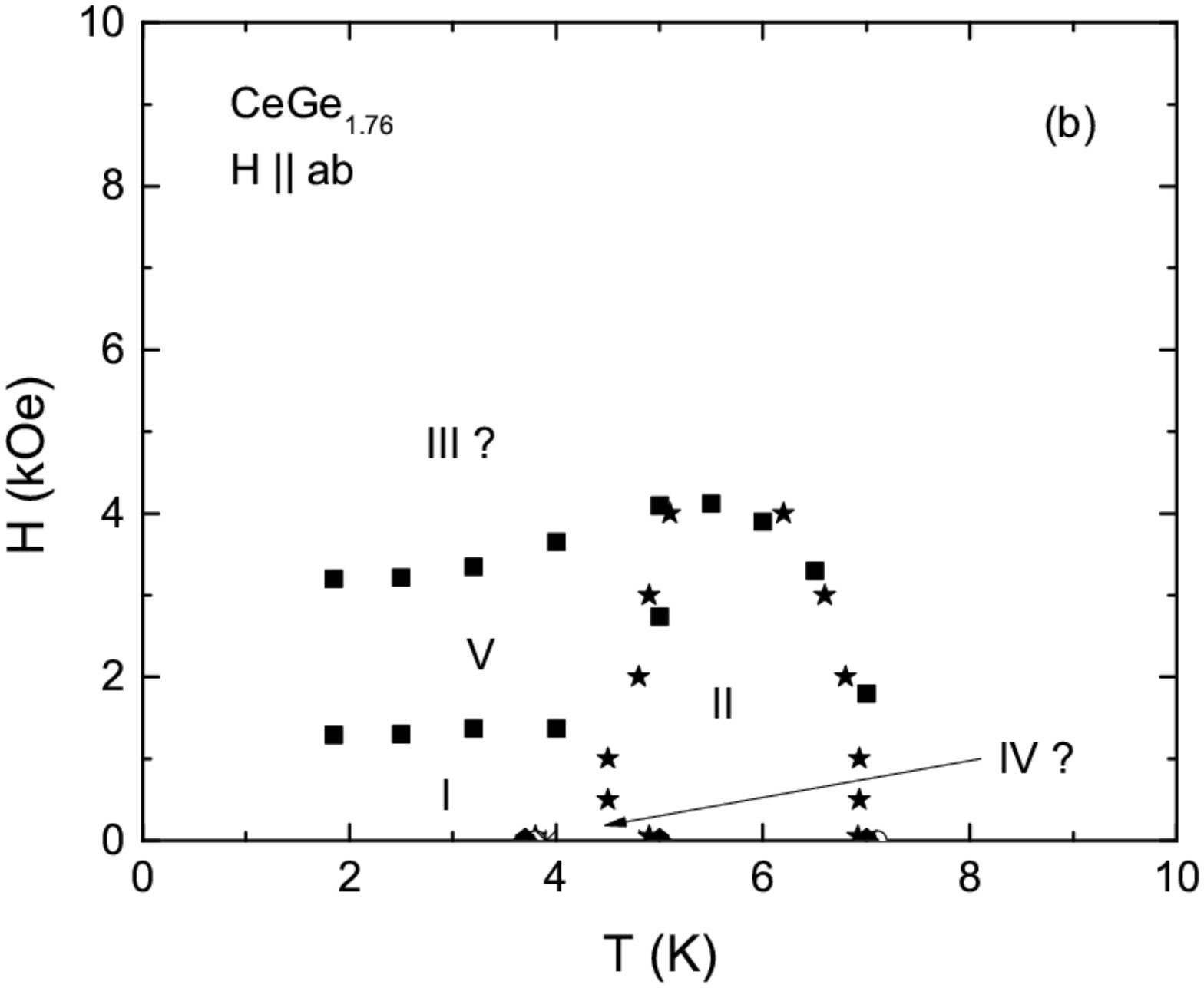}
\end{center}
\caption{Tentative $H - T$ phase diagrams for (a) $H \|  c$, and (b) $H \|  ab$. Symbols: squares - from $M(H)$ data, stars - from $M(T)$ data. For $H = 0$ data from resistivity, thermoelectric power, thermal expansion and heat capacity are added.  Roman numbers mark different magnetic phases} 
\label{F11}
\end{figure}

Additional features, corresponding to metamagnetic transitions, are observed in the anisotropic $M(H)$ data, taken up to higher, 10 kOe, field at different temperatures (Fig. \ref{F9}). Magnetization data between 10 kOe and 55 kOe for both orientations (not shown) are monotonic, with no additional features. The saturated magnetization in both directions is below $1 \mu_B$. The magnetization isotherms measured at different temperatures and low temperature, temperature-dependent magnetization measured at different fields (Fig. \ref{F10}) allow for construction an anisotropic, $H - T$, phase diagrams for CeGe$_{1.76}$ (Fig. \ref{F11}). The points obtained from $M(T)$ and $M(H)$ data are consistent with each other. For $H = 0$  the points obtained from resistivity, thermoelectric power, thermal expansion and heat capacity (as described in the text above) are added. Whereas the $H \| c$ direction is unique, the phase diagram for $H \| ab$ is shown for some guidance only, since twinning and exact orientation might affect the number and the exact position of the phase lines. Phase I is the phase with a ferromagnetic component along the $c$ - axis, phase II is a higher temperature antiferromagnetic phase. Phase IV, if indeed exists, is located $T \sim 4$ K and $T \sim 5$ K in zero field, and probably disappears by 0.5 - 1 kOe.  Phase V for $H \| ab$ is a metamagnetic phase.The phase marked as III appears to be the phase corresponding to saturated paramagnetic state. The magnetic field scales for CeGe$_{1.76}$ are rather small, not exceding 10 kOe, suggesting certain fragility of the magnetic order in this material.

\section{Summary}
We presented results of a comprehensive study of thermodynamic and transport measurements on CeGe$_{1.76}$ single crystals. Although formally it is an extreme member of the CeGe$_{2-x}$ family, the WDS data are consistent with  CeGe$_{1.75}$ or Ce$_4$Ge$_7$ and both resistivity data and magnetization loop data are consistent with a well ordered compound. This material has an enhanced electronic specific heat coefficient and two magnetic ordering transitions, at $\approx 7$ K and at $\approx 4$ K, and possibly a third one, at $T \approx 5$ K that is observed as a small anomaly in a number of measurements. The ground state characterized by a possibly compex magnetic structure with a ferromagnetic component along the $c$ - axis corresponding to a magnetic moment of $\approx 0.22 \mu_B$ per Ce. The response  of the  CeGe$_{1.76}$ to the applied magnetic field is anisotropic, with metamagnetism (spin reorientation) observed in both directions. This simple binary material is a good candidate for neutron scattering studies as well as for studies of the evolution of its magnetism under pressure. To some extent, CeGe$_{1.76}$ has similarities to other Ce-based, reduced moment ferromagnets, like CeAgSb$_2$ \cite{mye99a} and CeRu$_2$Ge$_2$ \cite{ray99a}, the studies of which yield  interesting results related to quantum phase transitions and strongly correlated behavior.

\ack
Help of W. E. Straszheim in WDS measurements on the sample is greatly appreciated. S.L.B. thanks A. I. Goldman for his abetment in finalyzing this manuscript. Work at the Ames Laboratory was supported by the US Department of Energy, Basic Energy Sciences, Division of Materials Sciences and Engineering under Contract No. DE-AC02-07CH11358. S.L.B. acknowledges partial support from the State of Iowa through Iowa State University.

\clearpage

\clearpage

\clearpage

\clearpage

\clearpage

\clearpage

\clearpage

\clearpage

\clearpage

\clearpage

\clearpage

\end{document}